\documentclass[
  ,draft            
  ]
  {aipproc}

\layoutstyle{8x11double}

\begin{document}

\title{Constructive role of non-adiabaticity for quantized charge pumping}

\classification{73.63.Kv, 73.21.La, 73.22.Dj, 73.23.Hk}
\keywords{single electrons pumps}

\author{B. Kaestner}{
	address={Physikalisch-Technische Bundesanstalt (PTB), Bundesallee 100, 38116 Braunschweig}
}

\author{C. Leicht}{
  address={Physikalisch-Technische Bundesanstalt (PTB), Bundesallee 100, 38116 Braunschweig}
}

\author{P. Mirovsky}{
  address={Physikalisch-Technische Bundesanstalt (PTB), Bundesallee 100, 38116 Braunschweig}
}

\author{V. Kashcheyevs}{
  address={Faculty of Computing, University of Latvia, Riga LV-1586, Latvia}
  ,altaddress={Faculty of Physics and Mathematics, University of Latvia, Riga LV-1002, Latvia}
}

\author{E. V. Kurganova}{
	address={Nijmegen High Field Magnet Laboratory, Toernooiveld 7, 6525 ED Nijmegen, The Netherlands}
}

\author{U. Zeitler}{
	address={Nijmegen High Field Magnet Laboratory, Toernooiveld 7, 6525 ED Nijmegen, The Netherlands}
}

\author{K. Pierz}{
  address={Physikalisch-Technische Bundesanstalt (PTB), Bundesallee 100, 38116 Braunschweig}
}

\author{H. W. Schumacher}{
  address={Physikalisch-Technische Bundesanstalt (PTB), Bundesallee 100, 38116 Braunschweig}
}

\begin{abstract}
We investigate a recently developed scheme for quantized charge pumping based on single-parameter modulation. The device was  realized in an AlGaAl-GaAs gated nanowire. It has been shown theoretically that non-adiabaticity is fundamentally required to realize single-parameter pumping, while in previous multi-parameter pumping schemes it caused unwanted and less controllable currents. In this paper we demonstrate experimentally the constructive and destructive role of non-adiabaticity by analysing 
the pumping current over a broad frequency range.
\end{abstract}

\maketitle


Pumping transport mechanisms have attracted much interest as an alternative means to generate charge and spin currents in the absence of a bias voltage. The pumping current results from periodic modulation of certain system parameters of a nanostructure connected to leads. Of particular interest has been the quantized regime when the current varies in steps of  $e \cdot f$  as a function of the system parameters, where $e$ is the electron charge and $f$ is the frequency of modulation.

Much effort has been devoted to the adiabatic regime, when the variation of the parameters is slow compared to relaxation times of the system, and current quantization has been achieved almost 20 years ago~\cite{pothier1PBI}. Recently, a scheme has been developed in which only one parameter is varied~\cite{blumenthal2007a, Kaestner2007c} and therefore non-adiabaticity is an essential requirement to achieve pumping~\cite{moskalets2002B}. In this regime the system is driven out of equilibrium, which was previously considered to counteract the quantized regime~\cite{Flensberg1999}. 
In the following we will demonstrate both, the destructive and constructive role of non-adiabaticity, depending on the pumping frequency $f$.

\begin{figure}
  \includegraphics[height=.3\textheight]{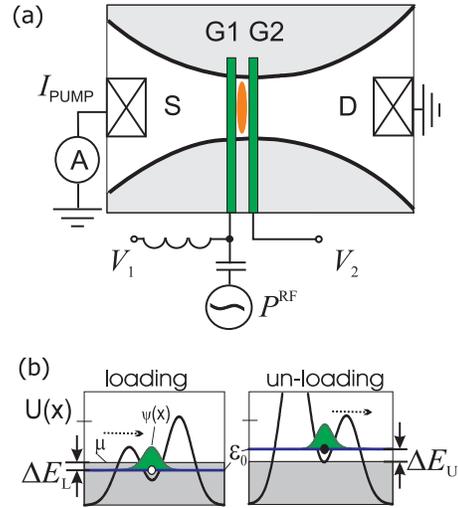}
  \caption{\label{figDevice} (a) Schematic of the device structure.
  (b) Schematic of the potential landscapes during loading and unloading as 
  generated by the device in (a).}
\end{figure}

\begin{figure}
  \includegraphics[]{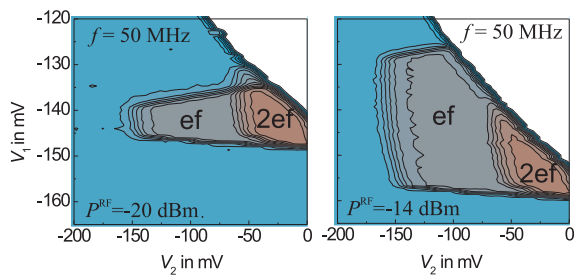}
  \caption{Pumping current as a function of $V_1$ and $V_2$ for 
  two different powers. Contours correspond to variation in current of 1.5$\,$pA 
  \label{figPower}}
\end{figure}


The device is similar to those used in~\cite{Kaestner2007c}. 
A schematic is shown in Fig.~\ref{figDevice}(a) and the corresponding 
potential landscape $U(x)$ along the channel in (b). 
It was realized in an AlGaAs/GaAs heterostructure. 
A 700$\,$nm wide constriction with a smooth curvature was generated inside the two-dimensional 
electron gas by etching the doped AlGaAs layer. 
The device was contacted at source (S) and drain (D) using an annealed layer of AuGeNi. 
The constriction is crossed by Ti-Au finger gates G1 and G2. A quantum dot (QD) with a
quasibound state $\psi$ is formed by applying sufficiently large negatice voltages $V_1$ and $V_2$
to G1 and G2, respectively. An additinoal sinusoidal signal of power $P^{RF}$ is coupled to G1. In this way the energy $\varepsilon_0$ of the quasibound state drops by $\Delta E_L$ below the chemical potential $\mu$ of the leads during the 
first half-cycle and can be loaded with an electron [see Fig.~\ref{figDevice}(b)]. During the second half-cycle, 
$\varepsilon_0$ is raised sufficiently fast by $\Delta E_U$ above $\mu$ and can be unloaded to
the right. Hence a current is driven through the sample. Details of the mechanism can be 
found in~\cite{Leicht2009}.


The pumped current $I$ as a function of $V_1$ and $V_2$ is shown in Fig.~\ref{figPower}
for two different powers. 
The measurements were performed at temperature $T=300\,$mK. 
Countours correpsond to variation in current of 1.5$\,$pA and
the plateaus to multiples of $e f$.
The plateaus are bound for more positive $V_1$ by insufficient unloading to drain
and for more negative $V_1$ by insufficient loading from source~\cite{Leicht2009}.
Variation in $P^{RF}$ shifts this boundary and allows calibration of the effective modulation
amplitude applied to the gate, as outlined in~\cite{kaestner2008}.
The step edges along $V_2$ are determined by escape of previously captured source-electrons 
back to source~\cite{kaestner2010a}. 
The white area in Fig.~\ref{figPower} corresponds to the region, where electrons 
may be loaded through the barrier at G2 and quantized pumping breaks down.

\begin{figure}
  \includegraphics[height=4.9cm]{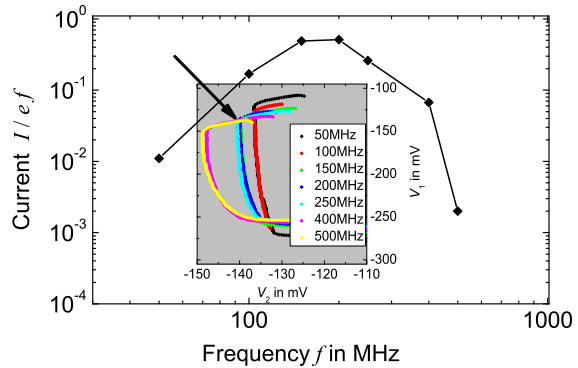}
  \caption{Current normalized by $e f$ as function of frequency. Inset shows 
  traced contours for the relevant range in $V_1$-$V_2$-space. The arrow 
  indicates the voltages where the current was measured. \label{figFrequenz}}
\end{figure}

In order to determine the role of non-adiabaticity in this pumping mechanism we analyse 
the frequency dependence at a point in $V_1$-$V_2$-space where both, the constructive and 
destructive nature becomes visible, namely for $V_1=-126\,$mV and $V_2=-140\,$mV.
The result is shown in Fig.~\ref{figFrequenz}. In order to extract the effects only due to
the $f$-dependence the voltage modulation amplitude at G1 needs to be kept constant.
To this end, the specific rf-power was determined for each frequency according 
to~\cite{kaestner2008}, and the $I = e f$ - plateau was analysed. The contour corresponding 
to $I = 0.5 e f$ is shown in the inset of Fig.~\ref{figFrequenz} for each frequency 
($f = 50, \ldots, 500\,$MHz). 
The contour was only traced for the relevant step-edge region.

The $I(f)$ dependence implies that
the average number of pumped electrons per cycle, $n_p = I/e f$, 
vanishes as $f$ is reduced, as expected when
only a single voltage parameter is modulated close to the adiabatic 
limit~\cite{moskalets2002B, Torres05, Kaestner2007c, Arrachea2005b}. 
From the inset of Fig~\ref{figFrequenz}
one can see that this corresponds to a shift in the left border of the $e f$-plateau 
toward positive $V_2$. 
Since this step-edge marks the transition where escape of previously
captured electrons back to source is prevented~\cite{Leicht2009, kaestner2010a} 
the constructive effect of non-adiabaticity 
becomes visible: only when $f$ is large enough, there will not be sufficient time for
escape and the electron will contribute to $I$.

Increasing $f$ beyond the optimal frequency range, $n_p$ reduces again as predicted in~\cite{Kaestner2007c}. From the inset of 
Fig~\ref{figFrequenz}
one can see that this corresponds to a shift in the upper border of the $e f$-plateau toward
negative $V_1$. Here the destructive nature of non-adiabaticity is illustrated:
since this step-edge corresponds to insufficient unloading to drain~\cite{Leicht2009} 
there is not sufficient time at such high frequencies to empty the dot. Consequently the electron
cannot contribute to the pumping current and $n_p$ drops. 

In general, driving a quantized current by a single modulation parameter, which is only possible in the non-adiabatic regime, is of fundamental importance in the development of a scalable quantum current standard~\cite{maisi2009}.

\emph{Acknowledgements:} This work has been supported by EURAMET joint research project with 
European Community's $7^{\mathrm{th}}$ Framework Programme, 
ERANET Plus under Grant Agreement No. 217257.
C.L. has been supported by International Graduate School of Metrology, Braunschweig.
Funding by EuroMagNET under the EU Contract No. RII3-CT-2004-
506239 is acknowledged.




\bibliographystyle{aipproc}   


\end{document}